\documentclass[runningheads]{llncs}
\usepackage{graphicx}
\usepackage{amsmath,amssymb}
\usepackage{physics}
\usepackage{bm}
\usepackage{float}
\usepackage{enumerate}
\usepackage{pgfplots}
\pgfplotsset{compat=1.9} 
\usepackage{tikz}
\usetikzlibrary{intersections, calc, arrows.meta,positioning}

\begin{document}
\title{Monotonicity of the Scalar Curvature of the Quantum Exponential Family for Transverse-Field Ising Chains}
\titlerunning{Monotonicity of the Scalar Curvature of the Quantum Exponential Family}
%
\author{Takemi Nakamura\orcidID{0000-0001-8797-2117}}
\authorrunning{T. Nakamura}
%
\institute{Nagoya University, Nagoya 464-8601, Japan \\
\email{nakamura.takemi.d7@s.mail.nagoya-u.ac.jp}}
\maketitle              
\begin{abstract}
The monotonicity of the scalar curvature of the state space equipped with the Bogoliubov-Kubo-Mori metric under more mixing a state is an important conjecture called the Petz conjecture.
From the standpoint of quantum statistical mechanics, the quantum exponential family, a special submanifold of the state space, is central rather than the full state space.
In this contribution, we investigate the monotonicity of the scalar curvature of the submanifold with respect to temperature for transverse-field Ising chains in various sizes and find that the monotonicity breaks down for the chains in finite sizes, whereas the monotonicity seems to hold if the chain is non-interacting or infinite-size.
Our results suggest that finite-size effects can appear in the curvature through monotonicity with respect to majorization.

\end{abstract}
\section{Introduction}
Investigating the state space of a physical system is often useful for considering its physical properties.
A physically important example of this is the Riemannian-geometric or information-geometric formulation of thermodynamics and statistical physics\cite{PhysRevLett.99.100602,ruppeiner1995riemannian}.
In information geometry\cite{amari2000methods}, the Gibbs distribution in classical statistical physics belongs to the exponential family, and a set of macroscopic variables or parameters in a statistical model can be viewed as a coordinate system of the statistical manifold.
The scalar curvature induced from the Fisher metric of the manifold is considered to be a physically significant quantity because it is related to thermal phase transitions and correlation\cite{ruppeiner2010thermodynamic,ruppeiner2014thermodynamic}.
Also in the quantum setting, a few works discuss the scalar curvature from the viewpoint of physics and suggest its physical significance\cite{nakamura2022scalar,PhysRevLett.99.100603}.

The Bogoliubov-Kubo-Mori (BKM) metric or the canonical correlation is one of the distinctive monotone metrics on the finite-dimensional quantum state space\cite{grasselli2001uniqueness,petz1994geometry,petz1993bogoliubov}.
The scalar curvature induced from the metric attracts interest in the mathematical context due to issues around the so-called Petz conjecture.
The conjecture states the monotonicity of the scalar curvature of the state space under majorization\cite{andai2003monotonicity,dittmann2000curvature,gibilisco2005monotonicity,petz1994geometry}.
In other words, it is conjectured that the scalar curvature shows entropy-like behavior.
However, the monotonicity of the scalar curvature of the quantum exponential family, a submanifold of the state space, is little discussed though the Gibbs state is central in quantum statistical mechanics.
The curvature tensor of a manifold is usually different from that of a submanifold of the manifold (e.g., a sphere in Euclidean space), hence one may wonder whether the scalar curvature of the quantum exponential family has the monotonicity and how it can be related to thermodynamic entropy.

The purpose of this study is to investigate the monotonicity property of the scalar curvature of the quantum exponential family with simple physical systems, transverse-field (TF) Ising chains in various sizes.
This is motivated by Ref. \cite{brody2003information}, which discusses finite-size effects on the scalar curvature.

This paper is organized into three sections: preliminaries, results, and discussion and conclusions.

\section{Preliminaries}
This section is devoted to introducing the concepts used here and fixing their notations.

\subsection{Scalar Curvature of the Quantum Exponential Family}

Let the Hilbert space of a quantum system be $\mathcal{H}$.
The quantum state of the system is generally described by a density operator on $\mathcal{H}$.
The set of all density operators is denoted by $\mathcal{S}(\mathcal{H}) := \{\hat{\rho} \, | \, \Tr\hat{\rho} = 1, \; \hat{\rho} \ge 0 \}$.

Our concern is a quantum statistical manifold, a parametric family of density operators
\begin{equation}
    \{ \hat{\rho}(x) \subset \mathcal{S}(\mathcal{H}) \; | \;  x \in \mathcal{X} \subset \mathbb{R}^n \},
\end{equation}
where $x=(x^1, x^2,..., x^n)$ denotes a set of the parameters forming a coordinate system whose domain denoted by $\mathcal{X}$ is a subspace of $\mathbb{R}^n$.
In this context, the Gibbs state at an inverse temperature\footnote{Here we set the Boltzmann constant to unity} $\beta = 1/T$ for a Hamiltonian $\hat{H}$, $e^{-\beta \hat{H}}/Z$, where $Z:= \Tr[e^{-\beta \hat{H}}]$ is the partition function, belongs to the following quantum exponential family
\begin{equation}\label{quantum exponential family}
    \hat{\rho}(\theta) = \exp[\theta^i \hat{\mathcal{O}}_i - \psi(\theta)]
\end{equation}
with the potential function
\begin{equation}
    \psi(\theta) := \ln \Tr[ \exp[\theta^i \hat{\mathcal{O}}_i] ] = \ln Z(\theta).
\end{equation}
Here $\theta = (\theta^1, \theta^2,..., \theta^n)$ and $\{ \hat{\mathcal{O}}_i \}_{i=1}^n$ denote a set of the natural parameters forming an e-affine coordinate system and a set of self-adjoint operators representing some physical observables, respectively.
We here assume that the self-adjoint operators are linearly independent of each other.
Note that this manifold is an $n$-dimensional submanifold of $\mathcal{S}(\mathcal{H})$

If density operators of a quantum system of interest are modeled in this way, the components of the BKM metric on the manifold can be given by the Hessian of the potential:
\begin{equation}\label{BKM metric for canonical parameter}
    g_{ij}(\theta) = \partial_i \partial_j \psi(\theta) ,
\end{equation}
where $\partial_i$ denotes $\partial/\partial\theta^i$\cite{ingarden1982information,janyszekRiemannianGeometryThermodynamics1989}.
The Christoffel symbols $\Gamma_{ijk}$ and the Riemannian curvature tensor $R_{ijkl}$ induced from this metric can be calculated as\cite{janyszekRiemannianGeometryThermodynamics1989}
\begin{align}
    \Gamma_{ijk}(\theta) &= \dfrac{1}{2} \psi_{ijk}(\theta) , \label{Christoffel symbol for Gibbs state} \\
    R_{ijkl}(\theta) &= \frac{1}{4} g^{ab} (\psi_{aik} \psi_{bjl} - \psi_{ail} \psi_{bjk} ) \label{def of scalar curvature},
\end{align}
where $\partial_i \partial_j \partial_k \psi(\theta)$ is abbreviated as $\psi_{ijk}(\theta)$.
Here we define the Riemannian curvature tensor as the curvature of a sphere becomes negative, in accordance with Ruppeiner\cite{ruppeiner1995riemannian,ruppeiner2010thermodynamic,ruppeiner2014thermodynamic}.
For the two-dimensional quantum exponential family, the scalar curvature can be calculated by this formula\cite{janyszekRiemannianGeometryThermodynamics1989}
\begin{equation}\label{scalar curvature in 2-dim for Gibbs state}
    R(\theta^1,\theta^2) 
    = \dfrac{2 R_{1212}}{\det g}
    = \dfrac{\mdet{\psi_{11} & \psi_{12} & \psi_{22} \\ \psi_{111} & \psi_{112} & \psi_{122} \\ \psi_{112} & \psi_{122} & \psi_{222} }}{2 \mdet{g_{11} & g_{12} \\ g_{21} & g_{22}}^2} .
\end{equation}

\subsection{TF Ising Chains}
The Hamiltonian of the TF Ising chain\cite{suzuki2012quantum} with nearest-neighbor interactions composed of $N$ qubits is given by
\begin{equation}
    \hat{H}_N = -J \sum_{i=1}^{N-1} \hat{\sigma}^z_i \hat{\sigma}^z_{i+1} - \Gamma \sum_{i=1}^{N} \hat{\sigma}^x_i,
\end{equation}
or when we impose the periodic boundary condition $\hat{\sigma}^z_{N+1} = \hat{\sigma}^z_{1}$ for $N \ge 3$, it can be also given by
\begin{equation}
    \hat{H}_N = -J \sum_{i=1}^{N} \hat{\sigma}^z_i \hat{\sigma}^z_{i+1} - \Gamma \sum_{i=1}^{N} \hat{\sigma}^x_i,
\end{equation}
where Pauli matrices $\hat{\sigma}^z_i$ and $\hat{\sigma}^x_i$ are represented as
\begin{equation}
    \hat{\sigma}^z_i := \mqty(1 & 0 \\ 0 & -1), \quad \hat{\sigma}^x_i := \mqty(0 & 1 \\ 1 & 0) .
\end{equation}
$J$ and $\Gamma$ represent an interaction and a transverse field, respectively.
The natural parameters are $(\theta, x) := (\beta J, \beta \Gamma)$.

\section{Numerical Results of Scalar Curvatures}
Once the potential for an equilibrium system with two parameters has been obtained, the scalar curvature for the system can be numerically computed using Eq. \eqref{scalar curvature in 2-dim for Gibbs state}.
In this section, we plot the graphs of scalar curvatures for TF Ising chains in sizes of $N=1, 2, 3, \infty$ and check if they are monotone with respect to temperature $T = 1/\beta$.
It should be noted that, if the Petz conjecture is true for the quantum exponential family, the scalar curvature here is expected to monotonically decrease as temperature increases.

\subsection{$N=1$ TF Ising Chain: Non-interacting Case}
This system consists of only one qubit, thus it can not have interactions.
Instead, we add contribution from a longitudinal field $h$ into the Hamiltonian:
\begin{equation}
    \hat{H}_1 = -h \hat{\sigma}^z - \Gamma \hat{\sigma}^x.
\end{equation}
This is often called the zero-dimensional TF Ising model and has correspondence with the one-dimensional classical Ising model.
We can regard this system as a non-interacting chain just by arranging independent qubits in a line because the two systems are essentially the same.
The potential for the model is
\begin{equation}
    \psi_1(z,x) = \ln(2\cosh r),
\end{equation}
where $(z,x):=(\beta h, \beta \Gamma)$ are the natural parameters, and $r:=\sqrt{z^2+x^2}$.
Therefore, the scalar curvature can be analytically obtained as
\begin{equation}\label{scalar curvature of 0D quantum Ising}
    R_{1} = \dfrac{2r-\tanh r}{2r^2\tanh r}\cosh^2r - \dfrac{1+\tanh^2r}{2\tanh^2r} .
\end{equation}

\begin{figure}
    \centering
    \includegraphics{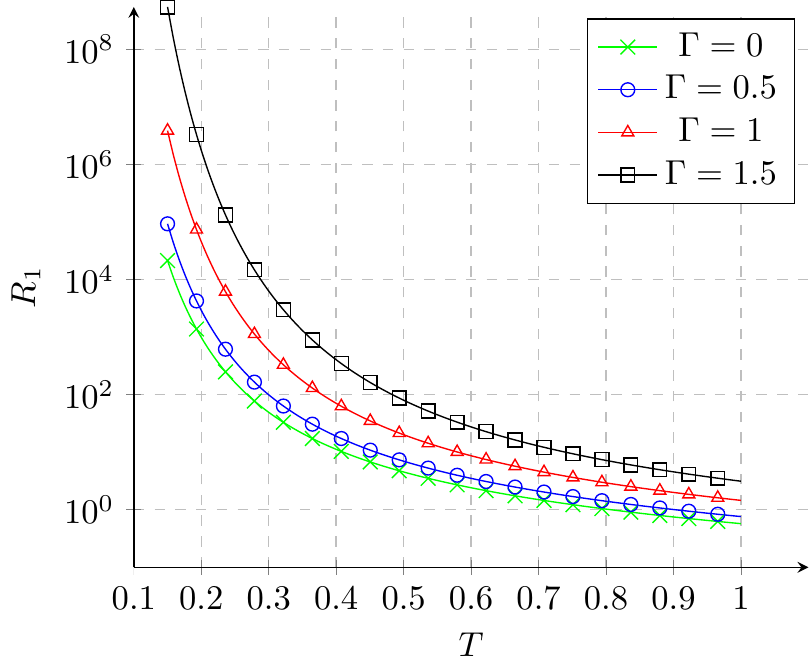}
    \caption{$R_1(T)$ with $J=1$ and different values of $\Gamma$ }
    \label{fig:R1}
\end{figure}

Fig. \ref{fig:R1} is a semi-log plot of this scalar curvature as a function of temperature.
As we can see, $R_1(T)$ decreases monotonically with respect to temperature.
In fact, we can check directly from Eq. \eqref{scalar curvature of 0D quantum Ising} that $R_1(T)$ is a monotonically decreasing function of temperature and hence $R_1(T) \ge 0$.

\subsection{$N=2$ and $N=3$ TF Ising Chains: Finite Size}
\begin{figure}
    \centering
    \includegraphics{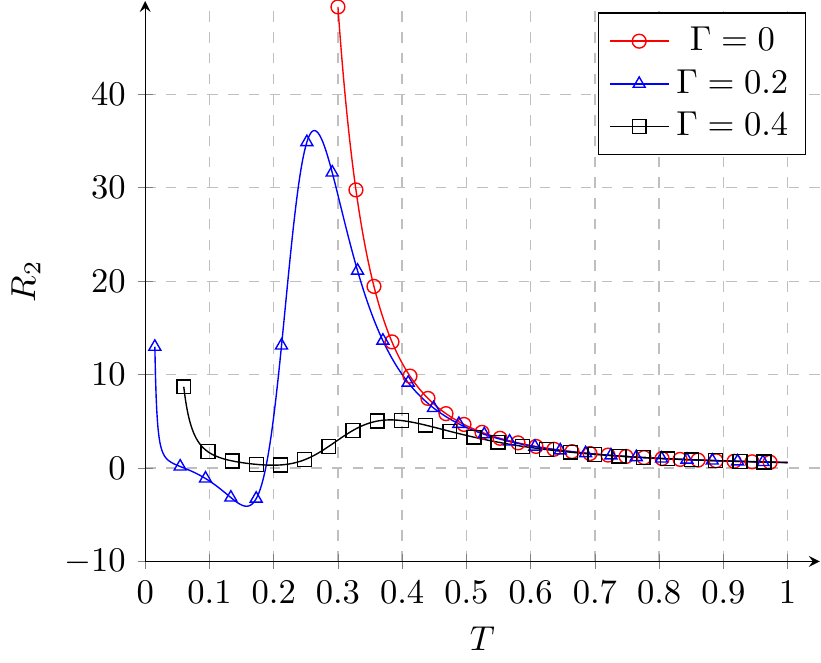}
    \caption{$R_2(T)$ with $J=1$ and different values of $\Gamma$}
    \label{fig:R2}
\end{figure}

\begin{figure}
    \centering
    \includegraphics{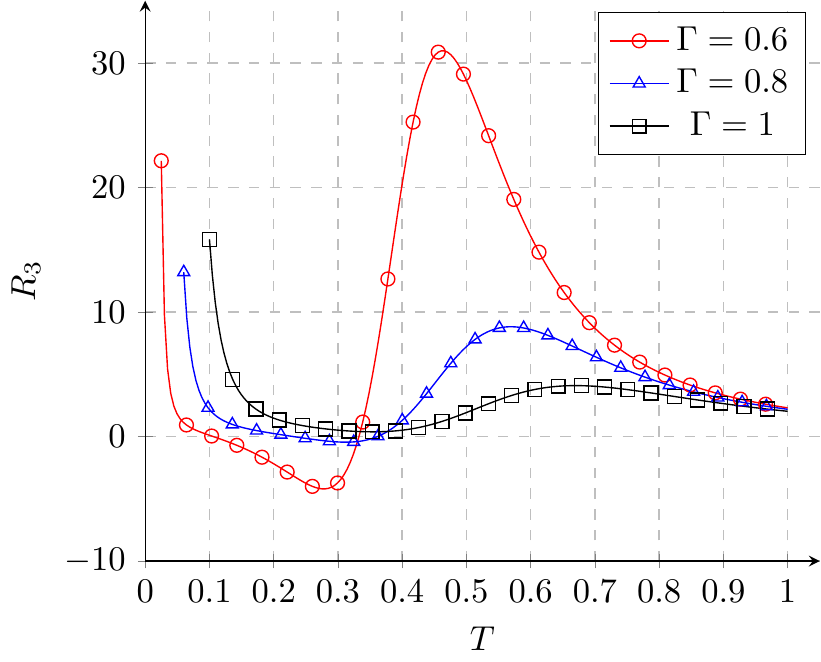}
    \caption{$R_3(T)$ with $J=1$ and different values of $\Gamma$}
    \label{fig:R3}
\end{figure}

Each potential is calculated as
\begin{equation}\label{potential of 1D TF Ising: N=2}
    \psi_{2}(\theta, x) = \ln \left[ 2 \cosh{\left(\theta \right)} + 2 \cosh{\left(\sqrt{\theta^{2} + 4 x^{2}} \right)} \right],
\end{equation}
\begin{align}\label{potential of 1D TF Ising: N=3}
    \psi_{3}(\theta, x) = \ln \left[ 2 e^{-\theta} \cosh(x) + 2 e^{\theta-x} \cosh(2\sqrt{\theta^2 + \theta x + x^2}) \right. \notag \\
     \left. + 2 e^{\theta+x} \cosh(2\sqrt{\theta^2 - \theta x + x^2}) \right],
\end{align}
where the periodic boundary condition $\hat{\sigma}^z_{4} = \hat{\sigma}^z_{1}$ is imposed for $N=3$.
The analytic expressions of the scalar curvatures $R_2, R_3$ are brutally complicated to compute, and we just give numerical plots: Fig. \ref{fig:R2} for $N=2$ and Fig. \ref{fig:R3} for $N=3$.

Surprisingly, these scalar curvatures wave and show three kinds of characteristic behaviors, resulting in non-monotonicity.
At higher temperatures, they seem to decrease monotonically and converge to zero.
In the middle, however, they moderately decrease and then increase, and they can have negative values during decreasing.
This behavior seems to vary depending on $\Gamma$.
Around zero temperature is also different: they seem to spike up to infinity as $T\to 0$.

\subsection{$N=\infty$ TF Ising Chain: Thermodynamic Limit}
In the thermodynamic limit $N=\infty$, only the potential per site or the potential density is meaningful because the potential itself has an extensive property.
In this case, we refer to the potential as the potential density.

\begin{figure}
    \centering
    \includegraphics{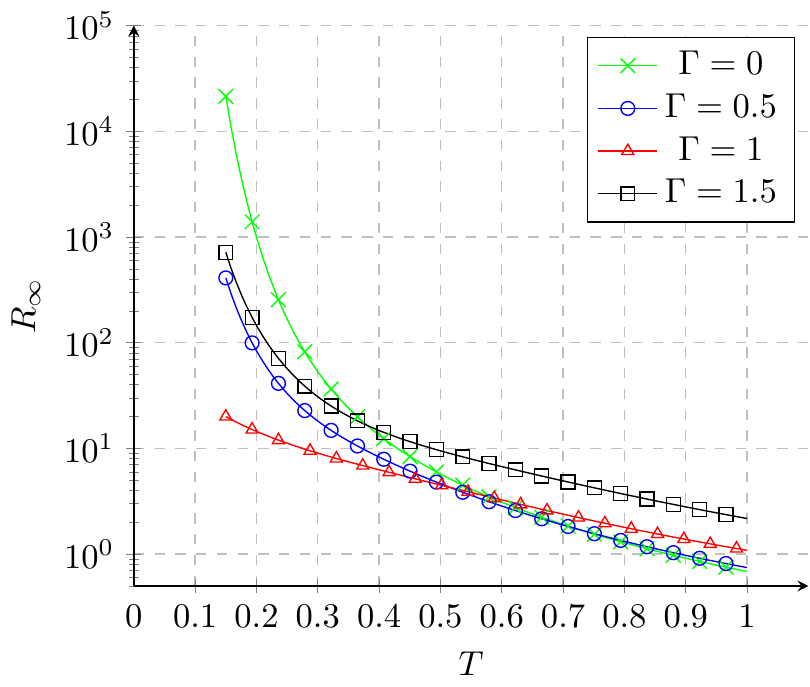}
    \caption{$R_{\infty}(T)$ with $J=1$ and different values of $\Gamma$}
    \label{fig:RInf}
\end{figure}

The potential can be obtained as
\begin{equation}\label{potential of 1D quantum Ising in the thermodynamic}
    \psi_{\infty}(\theta,x) := \dfrac{1}{N} \ln Z = \int_{0}^{\pi} \dfrac{\mathrm{d}k}{\pi} \ln(2\cosh f(k;\theta,x)) ,
\end{equation}
where $\displaystyle f(k;\theta,x) := \sqrt{\theta^2+x^2+2\theta x\cos k}$.
Numerical plots of the scalar curvature $R_{\infty}(T)$ for different $\Gamma$ are shown in Fig. \ref{fig:RInf}.
Unlike $R_2(T)$ and $R_3(T)$, $R_{\infty}(T)$ decreases monotonically, which confirms the monotonicity property.
In addition, these exponential behaviors are similar to $R_1(T)$ whereas $\Gamma$ dependence seems a little different.
Note that $\Gamma=J$ is the quantum phase transition point, hence $R_{\infty}(T)$ may expect to follow a power law, which is also confirmed in Fig. \ref{fig:RInf}.

\section{Discussion and Conclusions}
We have studied numerically the scalar curvature of the quantum exponential family for TF Ising chains in various sizes of $N=1$ (this case can be regarded as a non-interacting chain), $2$, $3$, and $\infty$ as a function of temperature, and checked whether it is monotone with respect to temperature.
We confirm the monotonicity for $N=1$ and $\infty$.
However, in finite-size cases $N=2$ and $3$ with an appropriate strength of a transverse field, the scalar curvature no longer possesses the monotonicity property and also can have negative values\footnote{Again the curvature tensor is defined here as that of a sphere becomes negative.}.
In any size, behavior at high temperatures looks alike.
Our results imply that
\begin{itemize}
    \item[(i)] $R(T\to \infty) = 0$ and $R(T\to 0) = \infty$ for an equilibrium system in any size,
    \item[(ii)] the Petz conjecture for the quantum exponential family is true for non-interacting or infinite-size (i.e., infinite-dimensional) quantum systems,
    \item[(iii)] finite-size effects can appear in the non-monotonicity of the scalar curvature with respect to majorization and may also cause negative curvature,
    \item[(iv)] we might be able to utilize the scalar curvature to judge whether an interacting quantum equilibrium system is in a finite size effectively.
\end{itemize}
Note that combining (i) and (ii) yields the non-negativity of the scalar curvature: $R(T) \ge 0$.
This is similar to thermodynamic entropy.
The physical interpretations of those behaviors of the scalar curvature require further investigation.

%
%
%
\bibliographystyle{splncs04}
\bibliography{mybibliography}

\end{document}